%% file: prd_for_CWR.tex
\documentclass[twocolumn,showpacs,aps,prd]{revtex4}

\usepackage{graphicx}
\usepackage{dcolumn}
\usepackage{amsmath}
\usepackage{epsfig}

\input babarsym.tex

\newcommand{\BABARPubYear}    {09}
\newcommand{\BABARPubNumber}  {025}

\newcommand{\SLACPubNumber} {13713}
\newcommand{\LANLNumber} {0911.1988}

\newcommand {\kpnunu}{\Kp \nunub}
\def\btoknunu	{\ensuremath{\Bu \to \Kp \nunub}\xspace}
\def\btodlnu	{\ensuremath{\Bub \to \Dz \ellm \nub X}\xspace}

\def\figurebox#1#2#3{%
    \def\arg{#3}%
    \ifx\arg\empty
    {\hfill\vbox{\hsize#2\hrule\hbox to #2{\vrule\hfill\vbox to #1{\hsize#2\vfill}\vrule}\hrule}\hfill}%
    \else
    {\hfill\epsfbox{#3}\hfill}%
    \fi}

\begin{document}

\preprint{\babar-PUB-\BABARPubYear/\BABARPubNumber} 
\preprint{SLAC-PUB-\SLACPubNumber} 

\begin{flushleft}
\babar-PUB-\BABARPubYear/\BABARPubNumber\\
SLAC-PUB-\SLACPubNumber\\
arXiv:\LANLNumber\ [hep-ex]\\[10mm]
\end{flushleft}

\title{
{\large \boldmath 
Search for the \btoknunu\space Decay Using Semi-Leptonic Tags} 
}

\input authors_jun2009

\date{\today}

\begin{abstract}
We present an update of the search for the flavor-changing neutral current \btoknunu decay using $351 \times 10^{6}$ \BB pairs collected at the \FourS resonance with the \babar\space detector at the SLAC PEP-II \B factory.  Due to the presence of two neutrinos in the final state, we require the reconstruction of the companion \B in the event through the decay channel \btodlnu.  We find 38 candidates in the data with an expected background of $31 \pm 12$.  This allows us to set an upper limit on the branching fraction for \btoknunu of $4.5\times 10^{-5}$ at 90\% confidence level. 
\end{abstract}


\maketitle
\section{Introduction}
\begin{figure}[h]
\begin{center}
  \includegraphics[height=3cm]{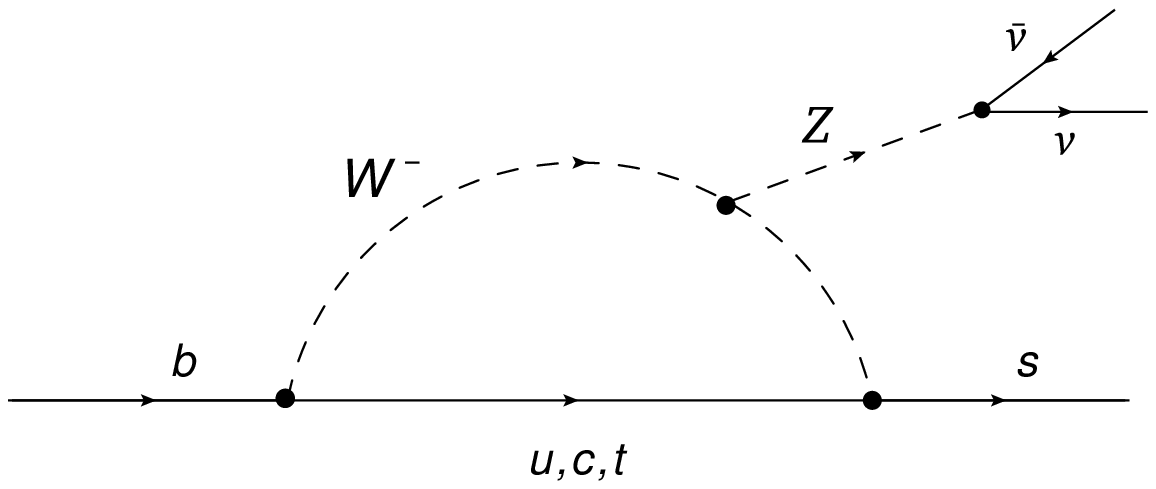}
  \includegraphics[height=4cm]{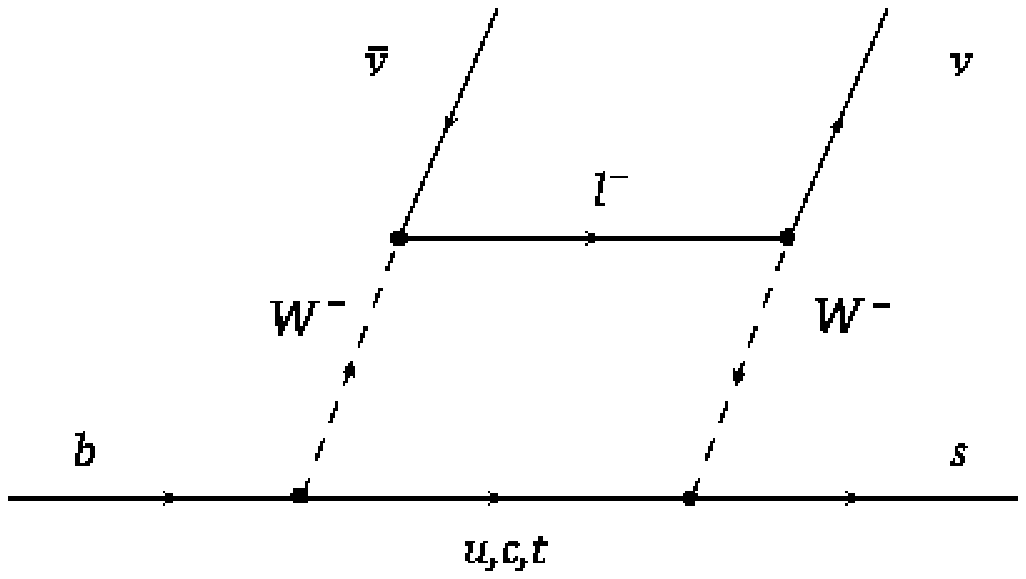}\\
   \caption{The $b\rightarrow s\nu\bar{\nu}$ transition proceeding through a penguin diagram (top) and a box diagram (bottom).\label{fig:feyn_diag}}
\end{center}
\end{figure}

Flavor-changing neutral-current transitions such as $b\rightarrow s\nu\bar{\nu}$ are absent at tree level in the Standard Model (SM) and occur only via electroweak penguin diagrams or one-loop box diagrams with virtual heavy particles in the loops, as shown in Fig.~\ref{fig:feyn_diag}.  Because such loop production processes are generally suppressed, the SM predicts $b\rightarrow s\nu\bar{\nu}$ transitions to be very rare.  We report herein the results of a search for the exclusive decay mode \btoknunu (charge conjugation is implied throughout this paper). The SM prediction of the branching fraction is $\BR(\btoknunu)=(3.8^{+1.2}_{-0.6})\times 10^{-6}$ \cite{Buchalla}, and the most stringent published upper limit at 90\% confidence level is  \BR$(\btoknunu)< 1.4\times 10^{-5}$ \cite{Belle Colab} by the Belle Collaboration with $535\times 10^{6}$ \BB pairs. 
This analysis serves as an independent measurement of the limit on this branching fraction.   

With current luminosity, we do not have the sensitivity to measure a branching fraction at the level of the SM; however several new physics models may increase the rate of $b\rightarrow s\nu\bar{\nu}$\space transitions. The Minimal Supersymmetric Standard Model with large $\tan\beta$ \cite{large tanb} leads to higher rates through chargino and/or charged Higgs contributions to the loop diagram.  ``Unparticle models'' \cite{unparticle} can give an observed enhancement because of the similarities in decay signatures between a $\btoknunu$ decay and a decay containing an unparticle.  Models with a single universal extra dimension \cite{extra dim} enhance the observed rate at lower values of $1/R$, where $R$ is the compactification radius of the extra dimension.  Also, models with light scalar dark matter candidates having \gevcc or sub-\gevcc\xspace masses \cite{wimps} may increase the observed rate because of the similarity in decay signatures between $\btoknunu$ and a decay containing two dark matter scalars.

Due to the presence of multiple neutrinos, the \btoknunu decay mode 
lacks the kinematic constraints that are usually exploited in \B decay 
searches at \B factories to reject both continuum (non-\BB) and \BB backgrounds. The strategy adopted for this analysis is to reconstruct an exclusive decay 
of the $B^{-}$ meson in the event, the ``tag \B,'' in one of several semileptonic decay modes.  All remaining charged and neutral particles in the event are examined under the assumption that they are products of the accompanying \B decay, the ``signal \B.''  We perform a multivariate analysis using a random forest classifier (explained below) to separate signal events from background events.  We keep the signal region of the classifier output blind to avoid experimenter bias.  The random forest classifier introduces very different systematic uncertainties than this collaboration's previous measurement \cite{Jack}.

\section{The \babar\space Detector and Dataset}
The data used in this analysis were collected with the \babar\ detector \cite{BABARNIM}
at the \pep2\ asymmetric \epem\ storage ring. The sample corresponds to an integrated luminosity of 319$\invfb$ at the \FourS resonance, and consists of about $351 \times 10^{6}$ \BB pairs. Charged-particle tracking and $dE/dx$ measurements for particle identification (PID) are provided by a five-layer double-sided silicon vertex tracker (SVT) and a 40-layer drift chamber (DCH) in a 1.5 T axial magnetic field. A ring imaging Cherenkov detector (DIRC) is used for $\pi-K$ discrimination.  The energies of neutral particles are measured by an electromagnetic calorimeter (EMC) consisting of 6580 CsI(Tl) crystals.  The magnetic flux return of the solenoid, instrumented with resistive plate chambers and limited streamer tubes, provides muon identification.

A {\sc Geant}4-based \cite{geant} Monte Carlo (MC) simulation is used to model the \babar\ detector response, taking into account the varying accelerator and detector conditions.  Dedicated signal and background MC samples are used to estimate the signal selection efficiency and determine the expected number of background events.   Simulation samples are used to model \BB events and continuum $\epem \to \uubar, \ddbar, \ssbar, \ccbar,$ and $\tautau$ events (background MC events).  A sample of $1.37\times 10^{6}$ events is simulated 
in which the 
$\Bp$ meson decays to \kpnunu, and the $\Bm$ meson decays to a mode with at least one lepton in the final state (signal MC events).

\section{Analysis Method}
\subsection{Tag \B Reconstruction}
The tag \B reconstruction combines a $D^{0}$ meson with a 
single identified charged lepton to form a $D^{0}\ell$ candidate.  The lepton candidate must have PID information consistent with an electron or a muon, have a minimum transverse momentum of 0.1\gevc, and have at least 20 hits in the DCH. 
The $D^{0}$ candidates are reconstructed in three decay modes: $K^{-}\pi^{+}$, $K^{-}\pi^{+}\pi^{-}\pi^{+}$, and $K^{-}\pi^{+}\pi^{0}$.  The charged pions from the $D^{0}$ decay must have a polar angle between 0.41 and 2.54 radians.
The $K^{-}$ candidate must fail $\pi^{-}$ PID requirements based on the candidate's measured $dE/dx$ for lower momentum candidates, or on the Cherenkov angle, number of photons, and track quality measured in the DIRC for higher momentum candidates.  The $\pi^{0}$ candidates are required to have a reconstructed mass between 0.115 $< m_{\pi^{0}} <$ 0.150\gevcc, and have an energy measured in the laboratory frame greater than 0.2\gev. 
The reconstructed $D^{0}$ mass, $m_{D^{0}}$, must be within 0.04\gevcc (0.07\gevcc) of the nominal $D^{0}$ mass for the channels without (with) a $\pi^{0}$ in the final state, and the center of mass momentum, $p_{D^0}$, must be greater than 0.5\gevc. The invariant mass, $m_{D^{0}\ell}$, of the $D^{0}\ell$ candidate must be greater than 3.0\gevcc.

Assuming that a neutrino is the only 
particle missing from a genuine $B^{-}\rightarrow D^{0}l^{-}\bar{\nu}$ decay, the cosine of the angle between the 
direction of the reconstructed tag \B and that of the $D^{0}l$ candidate, 
described by the four vector ($E_{D^{0}l},\textbf{p}_{D^{0}l}$), is given by

\begin{equation}\label{e:cosby}
\cos \theta_{B,D^{0}\ell}=\frac{2E_{B} E_{D^{0}\ell}-m_{B}^{2}-m_{D^{0}\ell}^{2}}{2|p_{D^{0}\ell}|\sqrt{E_{B}^{2}-m_{B}^{2}}},
\end{equation}

\noindent where $m_{B}$ is the nominal \B meson mass and $E_{B}$ and $\sqrt{E_{B}^{2}-m_{B}^{2}}$ are the expected \B meson energy and momentum, respectively, fixed by the energies of the beams and evaluated in the center of mass frame.  We retain events in the interval $-2.5<\cos \theta_{B,D^{0}\ell}<$1.1. These bounds are outside the allowed physical region to maintain efficiency for $B^{-}\rightarrow D^{*0}\ellm\bar{\nu}$ decays in which a $\pi^{0}$ or photon has not been reconstructed as part of the $D^{0}\ell$ combination and to account for resolution effects. If more than one $D^{0}\ell$ candidate is reconstructed in a given event, the one with the smallest $|\cos \theta_{B,D^{0}\ell}|$ is retained.  The $m_{D^{0}}$ distribution remains unbiased by this method of choosing the best candidate, allowing us to later use it for background estimations (SEC.~\ref{sec:background est}).

\subsection{Signal Event Selection}
Events containing a reconstructed tag \B are examined for evidence of a \btoknunu decay.  We require that the number of charged tracks remaining after the tag \B has been reconstructed is less than 4, and that the missing energy in the event is greater than 2.5 \gev.  The signal $K^{+}$ candidate must satisfy PID criteria, have a polar angle between 0.469 and 2.457 radians (the angular acceptance of the DIRC), and have a charge opposite that of the lepton in the tagged \B decay.  After applying these preliminary requirements, the MC samples are reweighted to reproduce the selection efficiencies determined on the data. We use a bagged decision tree multivariate classifier (random forest classifier \cite{breiman}), available in the StatPatternRecognition software package \cite{narsky}, to discriminate signal events from background. This method, a powerful alternative to ``rectangular cuts,'' separates events of different categories by training many decision trees (sequential partitioning of data into subsets of similar characteristics starting from a root node) \cite{breiman_book}. We choose to use a random forest classifier instead of another multivariate classifier because of its stability with higher dimensionality (more input variables), its training stability (the performance is less likely to diminish with continued training), and its insensitivity to input variables with weak discriminating power.

\begin{figure*}
\begin{center}
  \includegraphics[height=6cm]{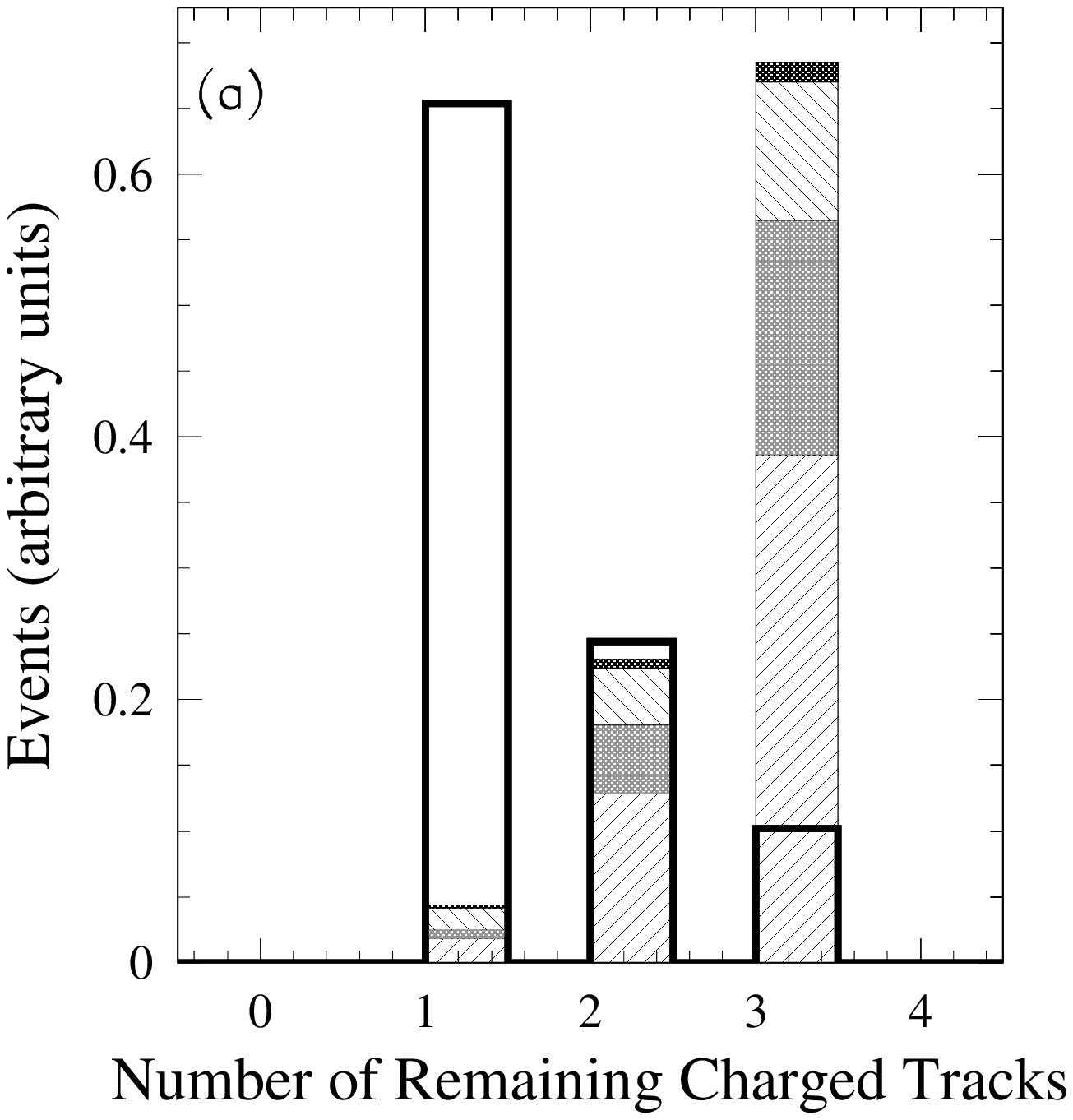}
  \includegraphics[height=6cm]{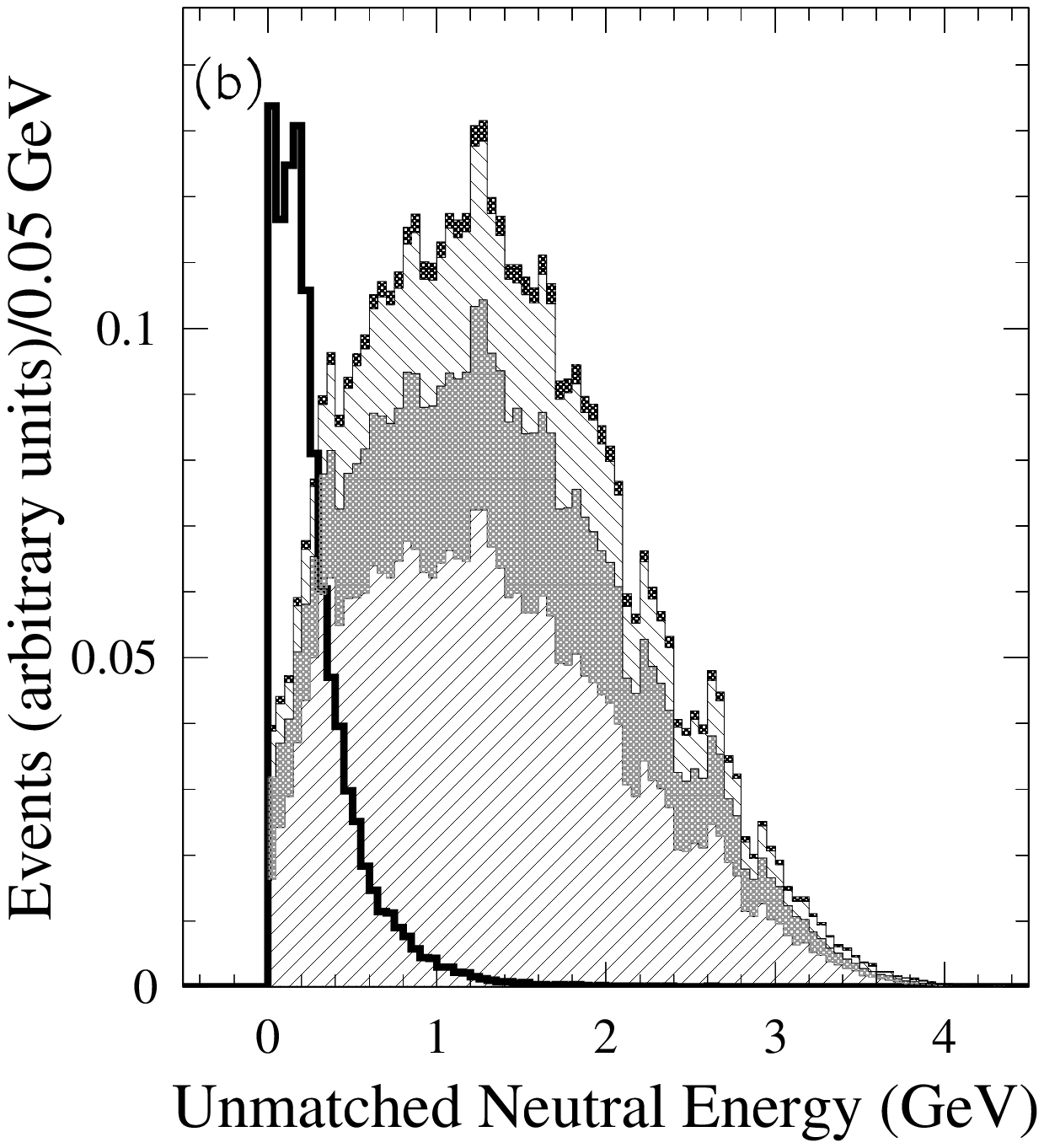}\\
  \includegraphics[height=6cm]{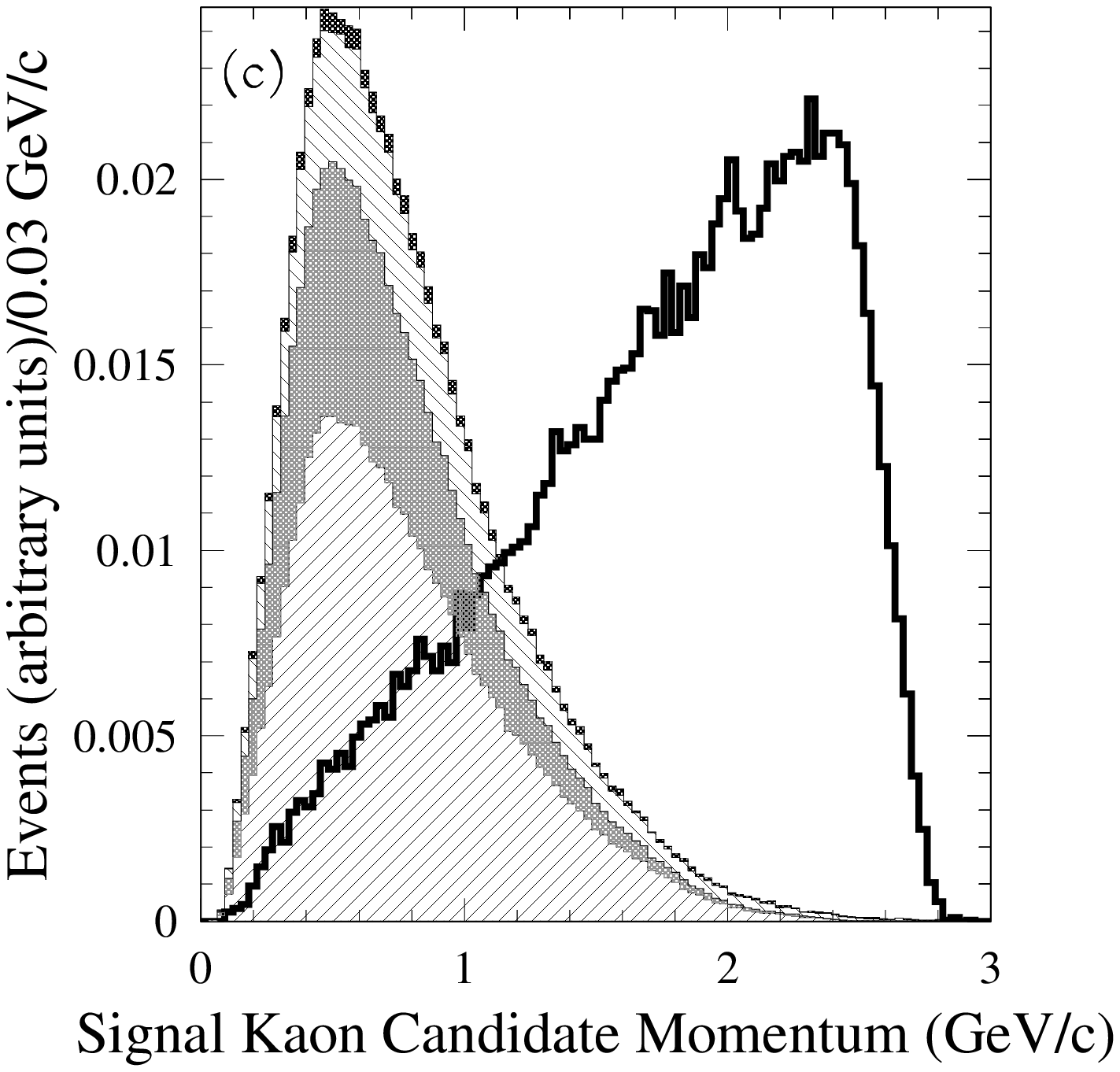}
  \includegraphics[height=6cm]{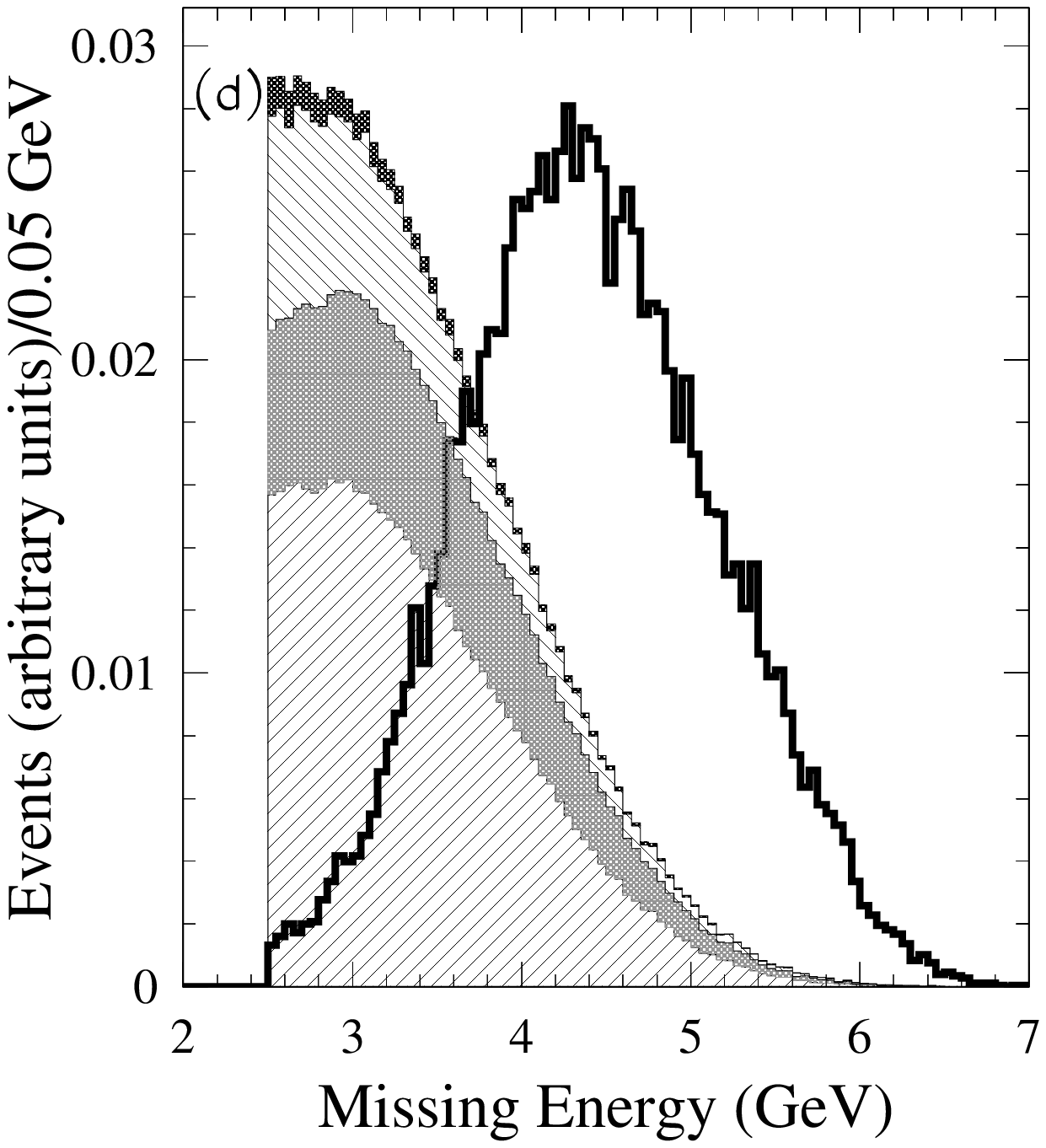}\\
  \caption{Several important variables that are used in the random forest classifier. (a) The number of charged tracks in the event not used for tag \B reconstruction, (b) the total energy of photons ($E_{\gamma} >$ 0.05\gev) not used in the tag \B reconstruction, (c) the momentum of the signal \B kaon candidate, and (d) the missing energy. These are shown for the signal MC events (bold black-outlined histogram)
and the different types of background; starting from the bottom of the stacked histogram, these are \BpBm (hatched), \BzBzb (light grey), 
\ccbar (hatched), and \uubar, \ssbar, \ddbar (black).  The signal 
MC distribution has  been normalized to unit area and the background classes 
have been normalized by an arbitrary constant, but reflect the relative amounts of each class as expected in the data.\label{fig:ntrkleft}}
\end{center}
\end{figure*}

We use the 19 variables listed in Table~\ref{tab:rf input variables} in the classifier, the most important of which are the number of charged tracks not used in the tag \B reconstruction, the total energy of signal side photons each with energy greater than 0.05 \gev (lower energy photons are not modeled as well in our MC events), the signal-kaon momentum, and the missing energy of the event.  Distributions of these variables (Fig.~\ref{fig:ntrkleft}) show a clear difference between signal events and the different types of background.

\begin{table}
\begin{footnotesize}
\begin{center}
\caption{\label{tab:rf input variables}Descriptions of the variables input to the random forest classifier.}
\begin{tabular}{l}
\hline
  \hline
\vspace{0.1mm}\\
   Tag \B Variables \\
\vspace{0.1mm}\\
  \hline
\vspace{0.1mm}\\

   1. Number of charged tracks used to reconstruct the \B \\
   2. Number of $\pi^{0}$'s in the $D^{0}$ decay mode \\
   3. The cosine of the angle between the thrust axis and the $z$-axis\\
\ \ \ \ in the center of mass \\
   4. Total momentum transverse to the $z$-axis\\
   5. Cosine of the angle of the momentum vector of the $D^{0}\ell$ \\
\ \ \ \ candidate to the $z$-axis in the center of mass\\
   6. Center of mass momentum of the lepton candidate \\
   7. Cosine of the angle between the tag side momentum of the \\
\ \ \ \ combined $D^{0}\ell$ and the momentum of the parent \B meson \\
\ \ \ \ in the center of mass

\vspace{0.1mm}\\
  \hline
\vspace{0.1mm}\\
   
   Signal \B Variables\\

\vspace{0.1mm}\\
  \hline
\vspace{0.1mm}\\

   8. Number of charged tracks remaining after reconstruction of \\
\ \ \ \ the tag side \\
   9. Total energy of photons with $E_{\gamma}>$ 50 MeV, after the \\
\ \ \ \ tag side reconstruction \\
   10. Momentum of the signal side kaon in the center of mass\\
   11. Number of photons with $E_{\gamma}>$ 50 MeV, remaining after the \\
\ \ \ \ \ tag side reconstruction \\
   12. Cosine of the angle between the signal side kaon and the \\
\ \ \ \ \ tag side lepton candidate in the center of mass\\
   13. Cosine of the angle between the signal side kaon and the \\
\ \ \ \ \ tag side $D$ candidate in the center of mass\\
   14. Cosine of the angle between the signal side kaon and the \\
\ \ \ \ \ tag side $D^{0}l$ candidate in the center of mass\\
   15. Number of $\pi^{0}$ candidates remaining after tag side \\
\ \ \ \ \ reconstruction \\
 
\vspace{0.1mm}\\
  \hline
\vspace{0.1mm}\\
   
   Event Variables\\

\vspace{0.1mm}\\
  \hline
\vspace{0.1mm}\\

   17. Amount of undetected energy\\
   18. Amount of undetected mass \\
   19. 2nd Fox-Wolfram moment \cite{FW}\\
\vspace{0.1mm}\\
    \hline
\hline
\end{tabular}
\end{center}
\end{footnotesize}
\end{table}

\begin{figure}[h]
\begin{center}
  \includegraphics[height=6cm]{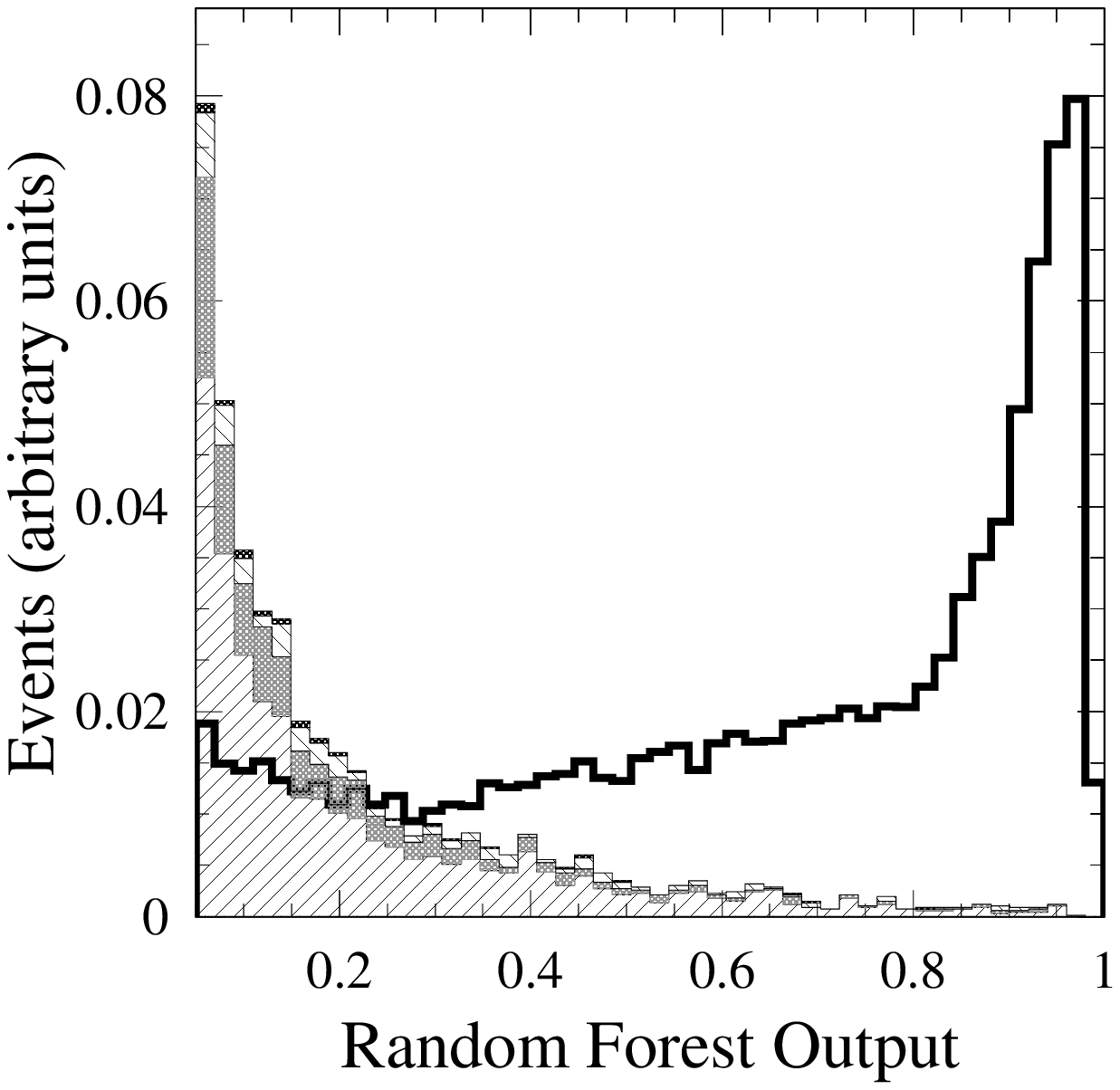}\\
  \caption{The output of the random forest classifier for the signal MC events (bold black-outlined histogram) 
and the different types of background; starting from the bottom of the stacked histogram these are \BpBm (hatched), \BzBzb (light grey), 
\ccbar (hatched), and \uubar, \ssbar, \ddbar  (black).  The signal 
MC distribution has  been normalized to unit area and the background classes 
have been normalized by an arbitrary constant, but reflect the relative amounts of each class as expected in the data.\label{fig:stacked}}
\end{center}
\end{figure}

 To avoid overtraining of the classifier, we separate the MC events equally between two samples. The classifier is trained on one sample, while the other is used to determine if the classifier is overtrained and what the signal and background efficiencies are when cutting on the classifier output.  We train the classifier by optimizing the Punzi figure of merit (FOM) \cite{Punzi}
\begin{equation}\label{e:punzi}
\frac{N_{S}}{0.5\cdot N_{\sigma}+\sqrt{N_{B}}},
\end{equation}
\noindent where $N_{S}$ is the expected number of signal events (assuming the SM branching fraction), $N_{B}$ is the number of expected background events, and $N_{\sigma}$ is the sigma level of significance, taken here to be 3. The output of the random forest classifier is shown in Fig.~\ref{fig:stacked} for MC simulated signal and background events.  Based on MC predictions, the multivariate classifier improves the FOM by 34\% over rectangular selection requirements.

We define the signal region by requiring the output of the multivariate classifier be greater than 0.82.  This selection is optimized using signal and background MC samples to maximize the FOM given in Eq.~\ref{e:punzi}. 
 When combined with the efficiency of reconstructing the tag \B, this gives a total signal efficiency of $\epsilon$ = (0.16$\pm$0.02)\%.  
The quoted error is the quadratic sum of the statistical, theoretical, and systematic uncertainties; we discuss the latter two below.  

\subsection{Background Estimation\label{sec:background est}}
\begin{figure}
\begin{center}
  \includegraphics[height=6cm]{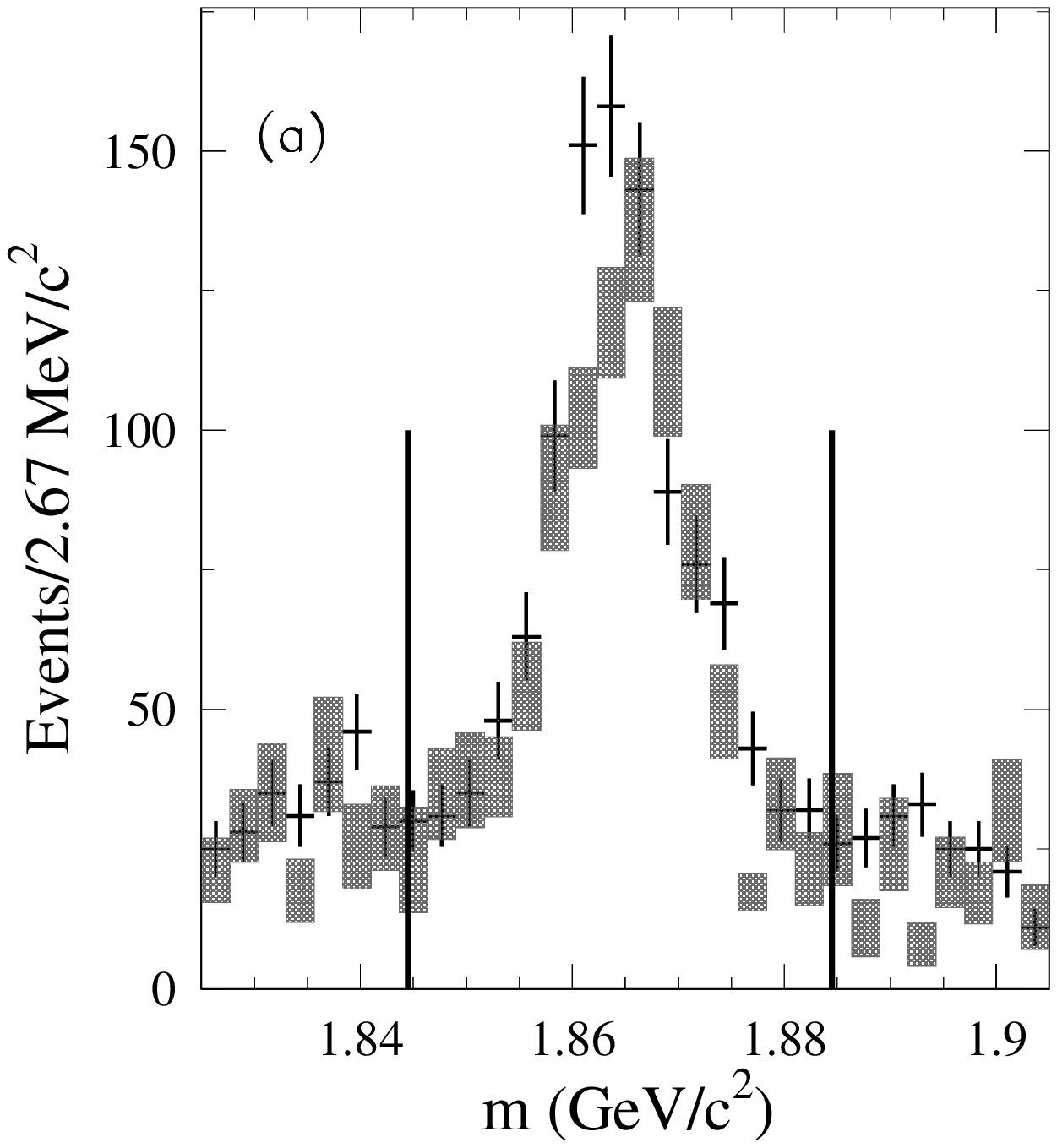}\\
  \includegraphics[height=6cm]{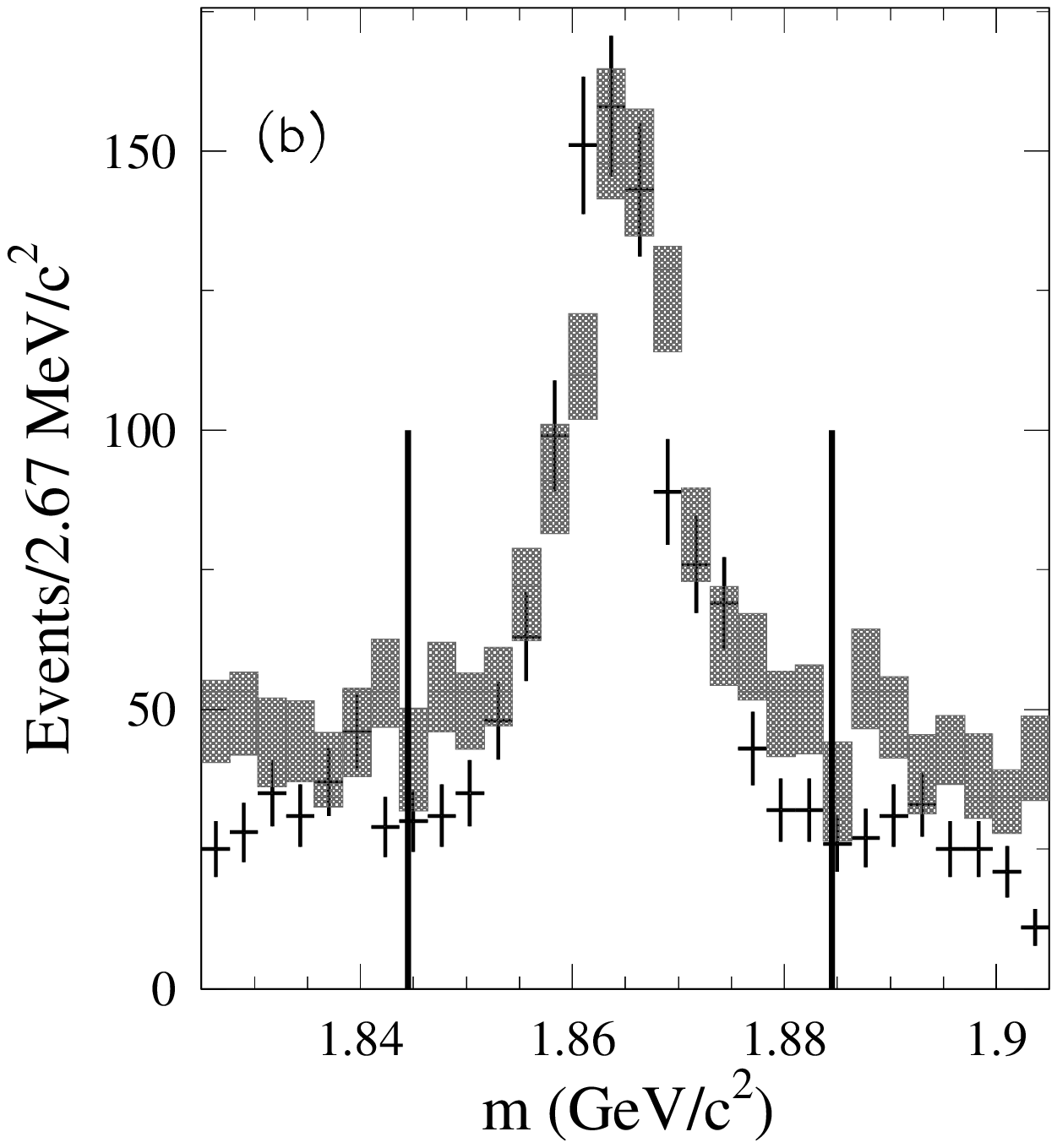}\\
  \includegraphics[height=6cm]{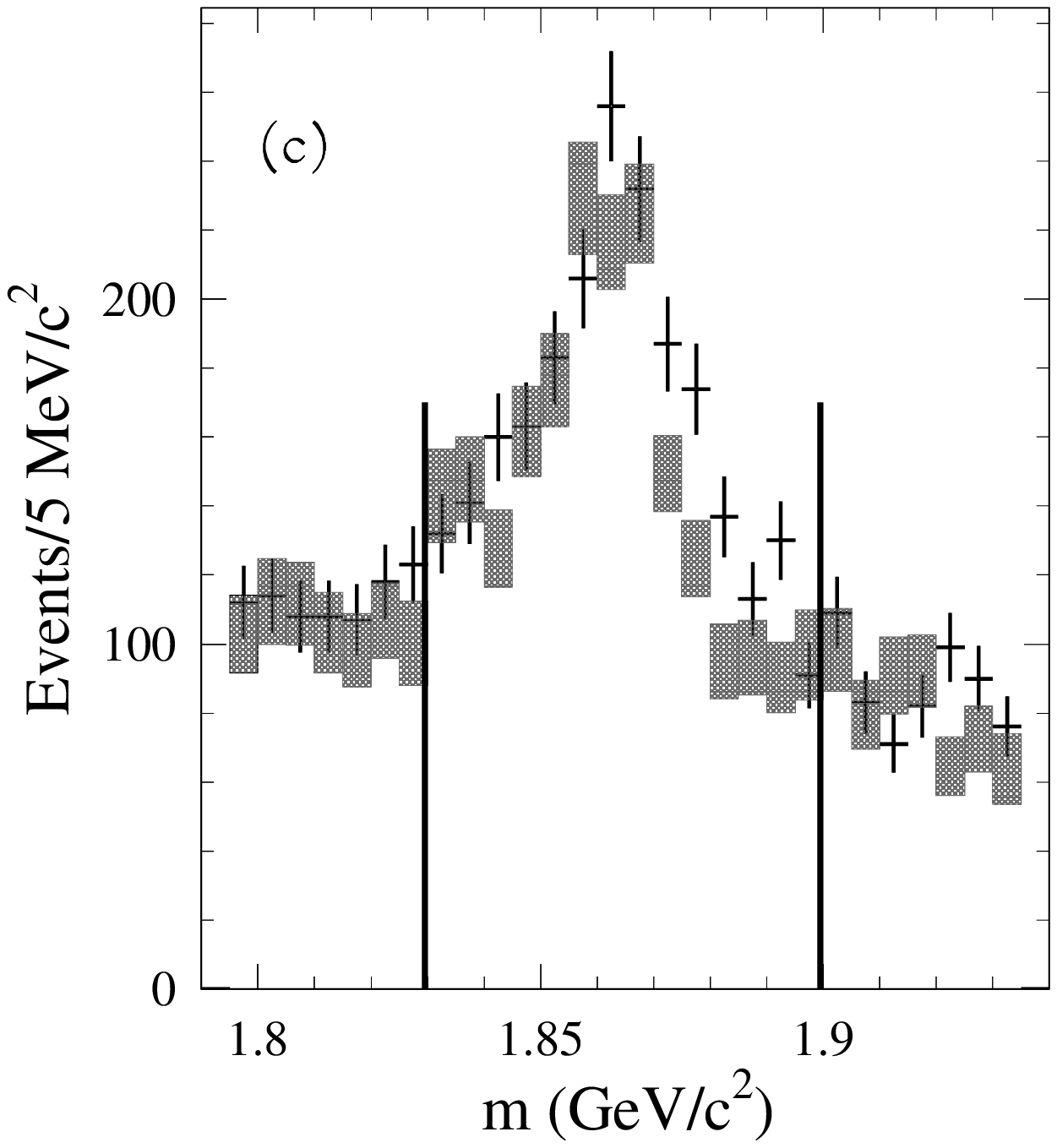}\\
\caption{A comparison of the invariant mass distributions of the $D^{0}$ candidates for the MC events and the data; the MC predictions are the grey rectangles (the height representing the uncertainty) and the data are the black points with error bars. The boundary between the signal and sideband regions is shown by the vertical black lines.  Plot (a) is the $K\pi$ mode, (b) is the $K\pi\pi\pi$ mode, and (c) is the $K\pi\pi^{0}$ mode. \label{fig:d0masses}}
\end{center}
\end{figure}

We divide the distribution of the invariant mass of the $D^{0}$ candidate into a high and low sideband and a signal region, defined in Table \ref{tab:d0sideband}.  This variable shows no correlation with the output of the random forest.  We identify two types of background in our signal region: combinatoric backgrounds, which are linear in the $m_{D^{0}}$ distribution shown in Fig.~\ref{fig:d0masses}, and peaking backgrounds, which correspond to true $D^{0}$ candidates and peak in the $m_{D^{0}}$ distribution. The combinatoric background level is estimated from the number of events in the $m_{D^{0}}$ sidebands that pass all cuts.  The level of combinatoric background expected in the signal region is 22 $\pm$ 5 events, where the uncertainty is statistical.

\begin{table}
\begin{center}
\caption{\label{tab:d0sideband} The mode-specific $m_{D^{0}}$ (in units of \gevcc) sideband definitions.  The boundaries for the third mode differ from those of the first two due to the presence of a $\pi^{0}$.}
\begin{tabular}{cccc}
\hline
\hline
\vspace*{0.1mm}\\
$D^{0}$ mode                & Lower Side        & Signal Region & Upper Side \\
\vspace*{0.1mm}\\
\hline
\vspace*{0.1mm}\\
$K\pi$, $K\pi\pi\pi$            & 1.8245-1.8445     & 1.8445-1.8845 & 1.8845-1.9045 \\
$K\pi\pi^{0}$       & 1.7945-1.8295     & 1.8295-1.8995 & 1.8995-1.9345 \\
\vspace*{0.1mm}\\
\hline
\hline
\end{tabular}
\end{center}
\end{table}

\begin{figure}
\begin{center}
  \includegraphics[height=6cm]{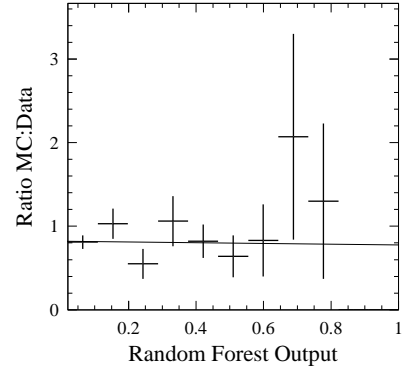}\\
  \caption{The ratio of the number of MC events to the number of data events in the $m_{D^{0}}$ signal region of each bin of the random forest classifier's output after background subtraction. The linear fit described in the text is also shown. \label{fig:rftrend}}
\end{center}
\end{figure}

To evaluate the peaking backgrounds, we start with the random forest output of
events in the $m_{D^{0}}$ signal region after subtracting the non-peaking part
using the $m_{D^{0}}$ sidebands. This distribution is produced for both MC and
data and their ratio is shown in Fig.~\ref{fig:rftrend}.  We fit a line to the points
in which the output of the classifier is less than 0.82 (the background region) and extrapolate it into the
signal region. The slope of the line is different from 0 and yields a multiplicative correction of 1.29 to the peaking component of our MC sample in the signal region that accounts for discrepancies between data and simulated events. The level of peaking background expected in the signal region is 9 $\pm$ 10 events, where the uncertainty is statistical. 

\section{Systematic Uncertainty Studies}

We vary the functions used to determine the estimated number of combinatoric and peaking background events in order to obtain the associated systematic uncertainties. This yields a systematic uncertainty of $\pm$1.9 events for the combinatoric background estimate and $\pm$3.2 events for the peaking background estimate. 
We associate an uncertainty with reweighting the MC samples after the preliminary requirements ($\pm$3.0 events), taken to be the difference between the number of events expected with and without this reweighting.

The systematic uncertainty associated with the requirement on the output of the random forest classifier is evaluated by selecting a double tag sample in both the data and the MC events in which both \B mesons decay semileptonically.  Using particle substitution on one of the two semileptonically-tagged \B mesons ($D^{0}\rightarrow K$, $l^{+}\rightarrow\nu$), we model the distributions of the variables that are included in the random forest classifier to resemble the signal MC events' distributions.  This second tagged \B meson serves as a control sample to estimate the difference in selection efficiency between MC events and data of the random forest classifier.  This difference between the double tag data and MC efficiencies (5.2\%) is assigned as the systematic uncertainty for how well the MC sample models the data.  An additional contribution of 9.3\% accounts for the difference between our control sample (double tag MC events) and our signal MC sample. Added in quadrature, the total uncertainty due to the random forest selection is 10.7\%.

Additional systematic uncertainties associated with the \btoknunu signal efficiency include uncertainties in the tagging efficiency (6.5\%), in the PID criteria used to identify the signal kaon (3.5\%), and in the tracking efficiency (0.5\%).  We evaluate the tagging uncertainty using the double tag sample, with both $D^{0}$s decaying to $K^{-}\pi^{+}$.  We take the ratio of the efficiency of finding both tags to the efficiency of finding one tag in both the MC sample and the data.  We then compare these ratios to determine the associated systematic uncertainty. 

The theoretical uncertainty on the $K^{+}$ momentum spectrum in \btoknunu
decays results in a 3.1\% uncertainty on the signal efficiency.  This
uncertainty is evaluated by comparing the efficiency obtained
using the kaon momentum spectrum given in Ref. \cite{Buchalla} and the efficiency
obtained when the sample is generated using a phase space model for
the decay.  These theoretical and systematic uncertainties on the signal efficiency are summarized in Table~\ref{tab:systematics}.

 \begin{table}
\begin{center}
\caption{\label{tab:systematics} Systematic uncertainties on signal efficiency (in \%).}
\begin{tabular}{cc}
\hline
\hline
\vspace*{0.1mm} \\
Systematic Uncertainties \\
\vspace*{0.1mm} \\
\hline
\vspace*{0.1mm} \\
Random Forest Selection  & 10.7\\
Tagging 		& 6.5\\
Kaon PID 		& 3.5\\
Tracking 		& 0.5\\
Kaon Momentum           & 3.1\\
\vspace*{0.1mm} \\
\hline
\vspace*{0.1mm} \\
Total                   & 13.4\\
\vspace*{0.1mm} \\
\hline
\hline
\end{tabular}
\end{center}
\end{table}

\section{Results}

\begin{figure}[h]
\begin{center}
  \includegraphics[height=6cm]{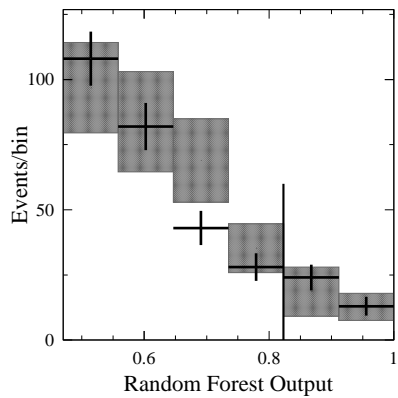}\\
  \caption{The distribution of the random forest classifier's output for data (crosses).  The expected range for the number of background events is shown as the grey boxes.  The vertical black bar shows the cut on the random forest classifier's output: the signal region is to the right.\label{fig:final_plot}}
\end{center}
\end{figure}

We expect 31 $\pm$ 12 background events in the signal region, and the SM predicts 2.2 signal events.  The signal efficiency is (0.16 $\pm$ 0.02)\%, which, using the Bayesian procedure described in \cite{Barlow}, yields an expected upper limit of 3.1$\times10^{-5}$ at 90\% confidence level. We observe 38 events in the signal region.  The distributions of the random forest classifier's output for the data and background Monte Carlo events are shown in Fig.~\ref{fig:final_plot}.  Because the number of events we see is consistent with the expected background, we interpret these results in the context of the SM, and set an upper limit on the branching fraction at \BR$(B^{+}\rightarrow K^{+}\nu\overline{\nu})< 4.5\times 10^{-5}$ at 90\% confidence level.  This is an improvement over \babar's previous search for this mode \cite{Jack}, which set an upper limit at \BR$(B^{+}\rightarrow K^{+}\nu\overline{\nu})< 5.2\times 10^{-5}$ using both semileptonic and hadronic tagged events, and is consistent with the recent, more stringent, limit by the Belle Collaboration \cite{Belle Colab}.

\input acknowledgements

\end{document}

%% file: authors_jun2009.tex
%
\author{B.~Aubert}
\author{Y.~Karyotakis}
\author{J.~P.~Lees}
\author{V.~Poireau}
\author{E.~Prencipe}
\author{X.~Prudent}
\author{V.~Tisserand}
\affiliation{Laboratoire d'Annecy-le-Vieux de Physique des Particules (LAPP), Universit\'e de Savoie, CNRS/IN2P3,  F-74941 Annecy-Le-Vieux, France}
\author{J.~Garra~Tico}
\author{E.~Grauges}
\affiliation{Universitat de Barcelona, Facultat de Fisica, Departament ECM, E-08028 Barcelona, Spain }
\author{M.~Martinelli$^{ab}$}
\author{A.~Palano$^{ab}$ }
\author{M.~Pappagallo$^{ab}$ }
\affiliation{INFN Sezione di Bari$^{a}$; Dipartimento di Fisica, Universit\`a di Bari$^{b}$, I-70126 Bari, Italy }
% ...
\collaboration{The \babar\ Collaboration}
\noaffiliation

%% file: acknowledgements.tex
We are grateful for the 
extraordinary contributions of our \pep2\ colleagues in
achieving the excellent luminosity and machine conditions
that have made this work possible.
The success of this project also relies critically on the 
expertise and dedication of the computing organizations that 
support \babar.
The collaborating institutions wish to thank 
SLAC for its support and the kind hospitality extended to them. 
This work is supported by the
US Department of Energy
and National Science Foundation, the
Natural Sciences and Engineering Research Council (Canada),
the Commissariat \`a l'Energie Atomique and
Institut National de Physique Nucl\'eaire et de Physique des Particules
(France), the
Bundesministerium f\"ur Bildung und Forschung and
Deutsche Forschungsgemeinschaft
(Germany), the
Istituto Nazionale di Fisica Nucleare (Italy),
the Foundation for Fundamental Research on Matter (The Netherlands),
the Research Council of Norway, the
Ministry of Education and Science of the Russian Federation, 
Ministerio de Educaci\'on y Ciencia (Spain), and the
Science and Technology Facilities Council (United Kingdom).
Individuals have received support from 
the Marie-Curie IEF program (European Union) and
the A. P. Sloan Foundation.